\documentclass{emulateapj}

\slugcomment{ApJ, Accepted}

\shorttitle{Constrains on the growth of the first supermassive black holes}
\shortauthors{Treister et al.}

\begin{document}

\title{New observational Constraints on the Growth of the First Supermassive Black Holes}

\author{E. Treister\altaffilmark{1}, K. Schawinski\altaffilmark{2}, M. Volonteri\altaffilmark{3} and P. Natarajan\altaffilmark{4,5,6}}

\altaffiltext{1}{Universidad de Concepci\'{o}n, Departamento de Astronom\'{\i}a, Casilla 160-C, Concepci\'{o}n, Chile; etreiste@astro-udec.cl}
\altaffiltext{2}{Institute for Astronomy, Department of Physics, ETH Zurich, Wolfgang-Pauli-Strasse 16, CH-8093 Zurich, Switzerland}
\altaffiltext{3}{Institut d' Astrophysique de Paris, 98bis Bd. Arago, 75014 Paris, France}
\altaffiltext{4}{Yale Center for Astronomy and Astrophysics, P.O. Box 208121, New Haven, CT 06520.}
\altaffiltext{5}{Department of Physics, Yale University, P.O. Box 208121, New Haven, CT 06520.}
\altaffiltext{6}{Department of Astronomy, Yale University, PO Box 208101, New Haven, CT 06520.}

\begin{abstract}
We constrain the total accreted mass density in supermassive black holes at $z$$>$6, inferred via the upper limit derived from the integrated X-ray emission from a sample of photometrically selected galaxy candidates. Studying galaxies obtained from the deepest Hubble Space Telescope images combined with the $Chandra$ 4 Msec observations of the $Chandra$ Deep Field South, we achieve the most restrictive constraints on total black hole growth in the early Universe. We estimate an accreted mass density $<\,$1000\,M$_\odot$Mpc$^{-3}$ at $z$$\sim$6, significantly lower than the previous predictions from some existing models of early black hole growth and earlier prior observations. These results place interesting constraints on early black growth and mass assembly by accretion and imply one or more of the following: (1) only a fraction of the luminous galaxies at this epoch contain active black holes; (2) most black hole growth at early epochs happens in dusty and/or less massive - as yet undetected - host galaxies; (3) there is a significant fraction of low-$z$ interlopers in the galaxy sample; (4) early black hole growth is radiatively inefficient, heavily obscured and/or is due to black hole mergers as opposed to accretion or (5) the bulk of the black hole growth occurs at late times. All of these possibilities have important implications for our understanding of high redshift seed formation models.
\end{abstract}


\section{Introduction}

One of the most challenging problems in astronomy today is understanding how and when the first supermassive 
black holes (SMBHs) in the Universe formed. It is widely believed that most of their growth happens primarily via accretion episodes \citep{soltan82}. The detection of luminous quasars at the earliest epochs implies that the rare behemoths with masses in excess 
of 10$^8$M$_\odot$ are already in place by $z$$\sim$7. The nature of the galaxies that host these is as yet unsettled 
due to the limitations of current observational technologies. Recently, it has been possible to detect the brightest most copiously star forming galaxies at these epochs via the photometric drop-out technique. A natural question is if these are the brightest and therefore the most massive galaxies at these epochs, then do they host the most massive black holes as the local demography of black holes suggests? We are now in a position to examine this issue with adequate data combining sources selected as part of the Hubble Ultra Deep Field (HUDF) and the Cosmic Assembly Near-IR Deep Extragalactic Legacy Survey (CANDELS). 

Nevertheless, deeply interconnected with this question are those of how the first black hole (BH) seeds form and when. Several possibilities for the formation of these seeds have been hypothesized \citep[see][for reviews]{volonteri10a,natarajan11}, however two remain the most accepted modes. The first one postulates that BH seeds result from the remnants of the first stars, the so-called population III generation stars \citep[e.g.,][]{abel00,madau01,bromm02}. Simulations suggest that these stars form from the collapse of primordial, metal-free, gas clouds at $z$$\sim$20, and have masses greater than 30~$M_\odot$ \citep{tan04}, implying very short lifetimes \citep{abel02}. Upon rapid exhaustion of fuel, these stars 
likely lead to the formation of seed BHs with masses $\sim$10-100~$M_\odot$ depending on the initial masses of the Pop III stars. The second possibility is a heuristic picture wherein early black hole seeds could form via direct gravitational collapse of gas-rich pre-galactic disks, leading to significantly more massive seed masses with $M_{\rm seeds}$$\sim$10$^{5}$$M_\odot$ \citep{loeb94,bromm03,begelman06,lodato06}.

Recent observations suggest that massive, a few M$\sim$10$^9$M$_\odot$, BHs were already in place by $z$$\sim$7 \citep[e.g.,][]{mortlock11}, i.e., $\sim$800 million years after the Big Bang. While these early massive BHs, which are detected as high-luminosity quasars, are very rare \citep{fan04, willott10a}, they do suggest that some fraction of BH seeds likely grow rapidly in the early Universe. These extreme sources, due to their extraordinary growth history, while individually interesting, have limited utility for probing the first BH growth episodes for the population as a whole. In fact, it is the lower mass, more common BHs, representative of the average population that are needed to constrain early growth and 
seed assembly scenarios. Directly examining such low mass -- and therefore low-luminosity -- BHs at $z$$\geq$6 is impossible with current observational instruments. The best we can do at the present time is therefore to stack data as the only way to access the earliest phases of black hole growth and discern the average properties of the population.

Observations at X-ray wavelengths are most suitable method for tracing BH growth as hard X-ray emission is the most reliable signpost for accretion. In this work, we study the X-ray properties of a sample of galaxies at $z$=6-8 selected based on the deepest observed-frame optical and near-IR images obtained with the Hubble Space Telescope (HST) to date from the HUDF and the CANDELS survey.  We take advantage of the 4 Msec Chandra observations of the Chandra Deep Field South (CDF-S) --- the deepest X-ray observation ever taken --- in order to constrain the integrated accreted BH mass density at $z$$>$6 and compare with existing models of BH formation and early growth. Throughout this letter, we assume a $\Lambda$CDM cosmology  with $h_0$=0.7, $\Omega_m$=0.27 and $\Omega_\Lambda$=0.73 \citep{hinshaw09}.

\section{Observational Results}

Thanks to the 4 Msec Chandra observations \citep{xue11}, the CDF-S is the field with the deepest currently-available X-ray observations. Our target sample of high-$z$ galaxy candidates was constructed using a 
combination of Lyman Break Galaxies (LBGs), selected using the optical and near-IR selection techniques described by e.g., \citet[][hereafter B06]{bouwens06}, and galaxies at $z$$>$6 based on their photometric redshifts 
obtained by performing spectral fitting. Using the deep HST observations available in this field, B06 reported the finding of 371 $z$$\sim$6 galaxy candidates, while later observations obtained using the WFC3 camera 
allowed for the detection of 66 at $z$$\sim$7 and 47 at $z$$\sim$8 \citep[][B11 hereafter]{bouwens11a}. More recently, \citet[][hereafter F12]{finkelstein12}, using the HST/WFC3 observations of the CANDELS fields obtained 
a sample of 223 galaxies at $z$$\sim$6, 80 at $z$$\sim$7 and 33 at $z$$\sim$8, all of them selected via photometric redshifts. The combination of these sources constitute the main sample for this work.

\begin{figure}
\begin{center}
\includegraphics[width=0.5\textwidth]{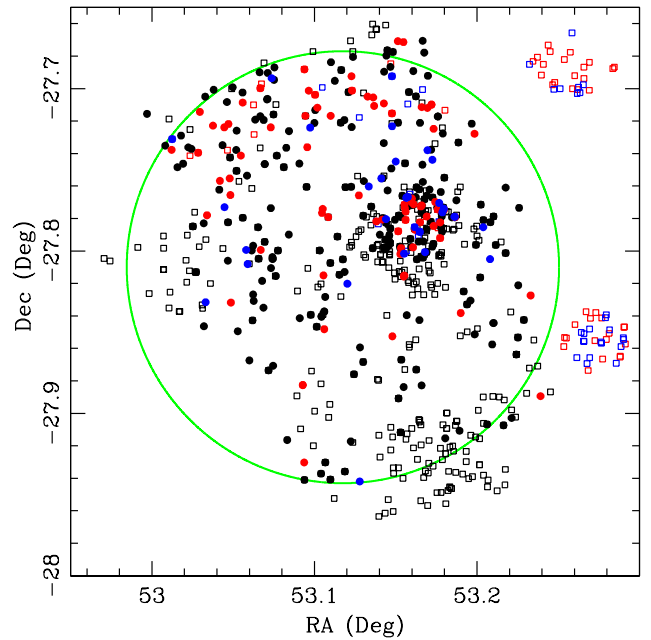}
\end{center}
\caption{Position in the sky of the sources considered in our work. Filled circles show the galaxy candidates from the sample of F12, while the
squares present the LBGs from the work of B06 ($z$$\sim$6) and B11 ($z$$\sim$7 and $z$$\sim$8). Sources at $z$$\sim$6 are shown in black, in
red the sources at $z$$\sim$7 and in blue at $z$$\sim$8. The field is centered in the aim point of the Chandra CDF-S observations, while the green
line shows a 8$'$ radius circle. Sources inside this circle were considered for X-ray stacking. The two groups clearly visible outside the green circle
correspond to the Hubble parallel fields.}
\label{sources_pos}
\end{figure}

As can be seen in Fig.~\ref{sources_pos}, the sources are not evenly distributed in the sky. The density is higher in the HUDF field, which is expected given the deeper optical and near-IR observations
available there. On average, sources in the HUDF are $\sim$3$'$ away from the Chandra pointing center. The solid area covered by the $z$$\sim$6 B06 sample is $\sim$160 arcmin$^2$, i.e. the GOODS-S region, 
which is almost completely included in the Chandra 8$'$ radius. At $z$$\sim$7 and $z$$\sim$8 sources are strongly clustered in the $\sim$5 arcmin$^2$ field. In the case of the F12 sample, sources are spread more 
evenly across the field, while a higher density is still observed in the HUDF region, due to the deeper data available there.

\begin{figure}
\begin{center}
\includegraphics[width=0.5\textwidth]{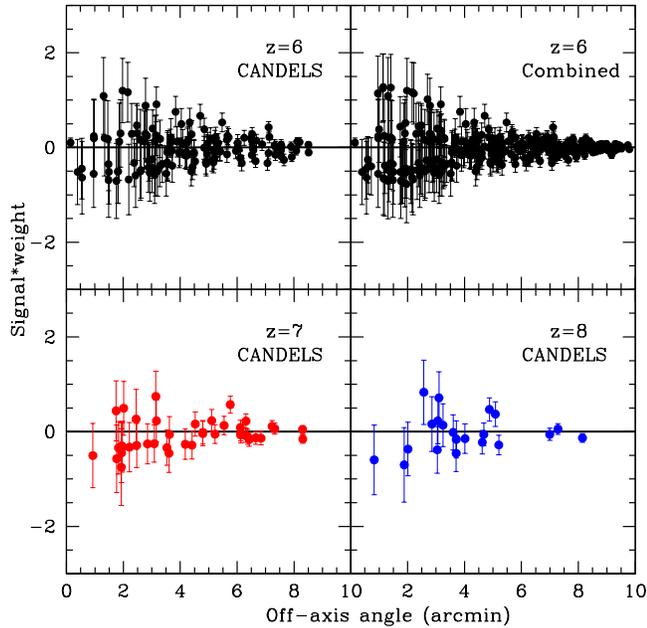}
\end{center}
\caption{Effective background-subtracted signal in the soft X-ray (0.5-2 keV) band measured using the procedure described by \citet{treister11} and the background subtraction algorithm of \citet{cowie12} for the 
sources in the F12 and the combined samples as a function of off-axis angle. The observed background-subtracted counts are weighted to account for the differences in exposure time, aperture corrections 
and sensitivity, as described by \citet{treister11}. Clearly, no source is individually detected beyond the 3-$\sigma$ level. Furthermore, the total signal is consistent with zero, thus confirming that no detection in the stacked 
samples is found either.}
\label{chandra_fluxes_soft}
\end{figure}

\begin{figure}
\begin{center}
\includegraphics[width=0.5\textwidth]{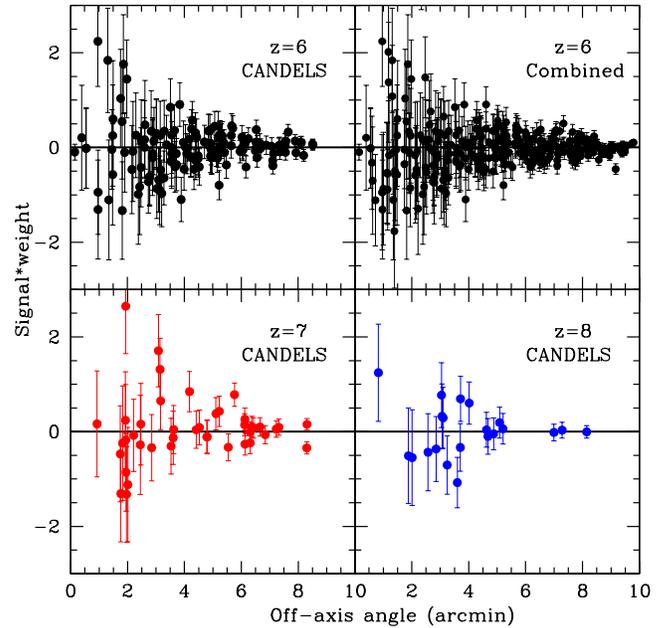}
\end{center}
\caption{Effective background-subtracted counts in the hard X-ray (2-8 keV) band measured using the procedure described by \citet{treister11} and the background subtraction algorithm of \citet{cowie12} for the 
sources in the F12 and the combined samples as a function of off-axis angle. Symbols are the same as in Fig.~\ref{chandra_fluxes_soft}. As in the previous case, no source is individually detected in this band 
beyond the 3-$\sigma$ level.}
\label{chandra_fluxes_hard}
\end{figure}

None of these high-$z$ candidates is detected individually in X-rays (using a 2\arcsec~search radius). Figures~\ref{chandra_fluxes_soft} and \ref{chandra_fluxes_hard} show the effective X-ray counts of each 
source --- further details can be found in \citet{treister11} and \citet{cowie12} --- for the sources in the B06, F12 and the combined samples. The X-ray properties of the sources at $z$$\sim$7 and $z$$\sim$8 in the B11 sample, not shown in these figures, are similar to those in the F12 sample at those redshifts. As can be clearly seen, no individual source is detected beyond the $\sim$3-$\sigma$ level in either the observed-frame  soft (0.5-2~keV) or hard (2-8 keV) Chandra bands. Therefore, we aim to detect X-ray emission from these sources using  stacking. Specifically, we followed the procedure of \citet{treister11}, modified to improve the background subtraction as described in \citet{cowie12}. The main goal of this procedure is to optimize the resulting signal-to-noise ratio (SNR), by introducing a variable aperture size as a function of the off-axis angle relative to the Chandra aimpoint. 

In order to maximize the SNR and avoid biasing our estimations of the local background we only considered sources closer than 8\arcmin~ from the Chandra aimpoint (03\fh32\fm28\fs06,-27\fdg48\farcm26\farcs4; \citealp{xue11}) that do not have a detected X-ray source at $<$15\arcsec. With these constraints, we stacked 223, 16 and 11 sources at $z$$\sim$6, $z$$\sim$7 and $z$$\sim$8 from the B06 and B11 samples respectively and 137, 40 and 20 sources from the sample of F12. Our results are presented in Table~\ref{stack_results}. As can be seen, there is no significant detection in any of the samples, thus contradicting the earlier results claimed by \citet{treister11}. This confirms that, as presented by \citet{willott11} and \citet{cowie12}, the results in \citet{treister11} likely arise from their background subtraction technique. 

As shown in Figure~\ref{sources_pos}, given that the depth of the optical and
near-IR coverage of the field studied here is not homogeneous, and
that we did not obtain a significant detection by stacking the entire
sample of galaxies at $z$$\sim$6, $z$$\sim$7 and $z$$\sim$8, we attempted to stack
the most luminous optical/near-IR galaxies, which can be detected
across the whole field. Specifically, we performed independent stacks
for different cuts in optical and near-IR fluxes. None of these
stacks yielded a significant detection either.

\begin{figure}
\begin{center}
\includegraphics[width=0.5\textwidth]{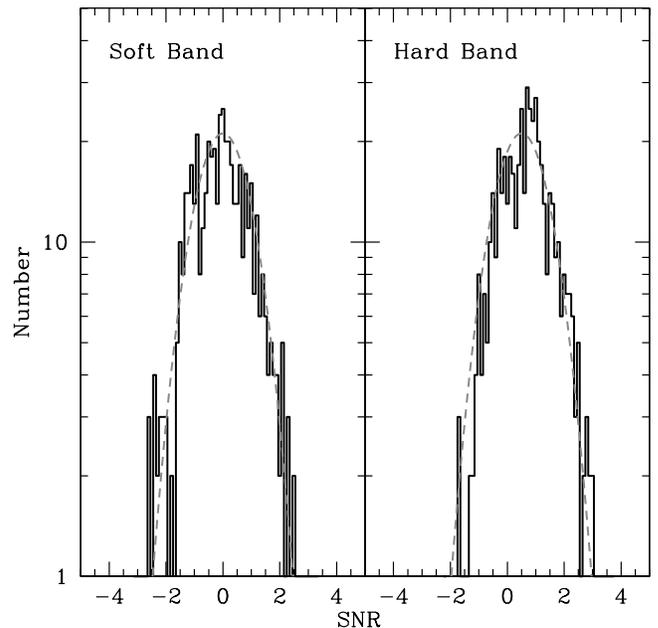}
\end{center}
\caption{Distribution of SNR obtained for 500 independent stacks, in which the position of each stacked source was shifted randomly in the soft (\textit{left panel}) and hard (\textit{right panel}) Chandra bands.
The \textit{gray dashed lines} in each panel show the best-fitting gaussian fits, with mean $\simeq$0 (0.5 for the hard band) and standard deviation $\simeq$1.}
\label{mc_dists}
\end{figure}

To establish the statistical significance of the non-detections in the entire sample and translate them into upper
limits on the stacked X-ray emission we perform independent Monte Carlo simulations for each of these 
samples. This is done by computing the obtained SNR in 500 stacks, in 
which the position of each stacked source was shifted randomly in right ascension and declination in the 5\arcsec-15\arcsec range. For both bands, the resulting distributions are well fitted by Gaussian functions with mean $\simeq$0 and standard deviation $\simeq$1, as expected. The SNR distributions obtained in our Monte Carlo simulations is shown in Figure~\ref{mc_dists}.

We then compute 3$\sigma$ upper limits for the X-ray emission from these samples. Focusing solely on the joint samples, in order to maximize the number of sources, we find for the $z$$\sim$6 sources an upper limit for the X-ray luminosity of 2.6$\times$10$^{41}$~erg~s$^{-1}$ in the soft band, $\sim$2$\times$ smaller than the upper limit reported by \citet{cowie12}. This is as expected due to the increased sample size providing a tighter upper limit. In the hard band, we compute an upper limit of 1.6$\times$10$^{42}$~erg~s$^{-1}$. Similarly, for the $z$$\sim$7 galaxies the upper limits are 6.8$\times$10$^{41}$~erg~s$^{-1}$ and 5.3$\times$10$^{42}$~erg~s$^{-1}$ in the soft and hard bands respectively. At $z$$\sim$8, these are 1.5$\times$10$^{42}$~erg~s$^{-1}$ and 9.8$\times$10$^{42}$~erg~s$^{-1}$.

Similar measurements were performed recently by \citet{fiore12} and \citet{basu-zych13} among others, finding consistent results. For example, \citet{basu-zych13} report upper limits in the rest-frame 2-10 keV
band of 4.2$\times$10$^{41}$~erg~s$^{-1}$ at $z$$\sim$6, 9.5$\times$10$^{41}$~erg~s$^{-1}$ at $z$$\sim$7 and 1.6$\times$10$^{42}$~erg~s$^{-1}$ at $z$$\sim$8. These are slightly higher but fully consistent with the
upper limits for the stacked X-ray luminosities reported here. The (small) differences can be due to the stacking technique used and the minor differences between the galaxy samples.

It is important to note that both the samples of B06 and F12 explicitly exclude point-like (spatially unresolved) sources. Most of the high-z galaxy candidates are spatially resolved, and so only only a small fraction of sources ($\sim$5\%) 
are removed by this requirement. While this is done to reduce the stellar contamination, it could represent a bias against unobscured Active Galactic Nuclei (AGN) in which the nuclear emission dominates the optical light. In order to test for the possible effects of
this criterion, we independently stack the sample of 11 point-like $i$-dropouts in the HDF-S reported by \citet{bouwens06}, in their Table 3. Consistent with the results obtained from the resolved sample, we do not find a significant detection in 
X-ray stacking in neither the soft nor the hard $Chandra$ bands. Therefore, we conclude that the exclusion of point-like sources from our main samples does not affect the results reported here.

\section{Discussion}

With the upper limits obtained above, we can derive the most stringent observational constrains to date on the early growth of the first SMBHs. We first note that the upper limits obtained here are lower than the standard threshold for AGN-dominated X-ray emission \citep[e.g.][]{szokoly04}, $\sim$10$^{42}$~erg~s$^{-1}$. Therefore, we can conclude that {\bf either no luminous AGN are present in any of the samples studied here or the occupation fraction of such AGN is so low that the signal gets diluted in the stacking procedure}. 

If we proceed under the assumption that the luminosity of these sources derives entirely from star formation (i.e. no AGN are present), as proposed previously by \citet{cowie12},  these upper limits can then in turn be used as an observational constraint on the average star formation rate (SFR). Taking into account the tight correlation between X-ray luminosity and SFR in absence of AGN emission measured by \citet{ranalli03} and more recently by \citet{lehmer10}, we can compute an upper limit for the average SFR in these samples. Since these are relatively low-mass galaxies, with typical stellar masses $<$10$^{10}$M$_\odot$ \citep{gonzalez11}, we can safely neglect the contribution of low mass X-ray binaries to the total X-ray luminosity. Therefore, we can assume that the X-ray luminosity is proportional to the star formation rate (SFR), with a relation given by $L_{HX}$=$\beta$~SFR, where $L_{HX}$ is the rest-frame 2-10 keV luminosity, and  $\beta$=1.62$\times$10$^{39}$erg~s$^{-1}$(\,M$_\odot$yr$^{-1}$)$^{-1}$ \citep{lehmer10}. While this was established from observations of $z$$\simeq$0 galaxies, no evolution in this correlation has been observed up to $z$$\sim$1.3 \citep{mineo12}. Assuming that we can extrapolate this relation to $z$$\sim$6 enables the translation of the upper limits of the X-ray luminosity obtained above to SFRs of $>$210\,M$_\odot$yr$^{-1}$ for the $z$$\sim$6 samples, $>$460\,M$_\odot$yr$^{-1}$ at $z$$\sim$7 and $>$1000\,M$_\odot$yr$^{-1}$ at $z$$\sim$8. In comparison, these upper limits on the SFRs are significantly higher, by an order of magnitude or more, than the estimated values for the SFR reported by \citet{gonzalez10,curtis-lake13} of $\sim$5-20M$_\odot$yr$^{-1}$ for individual galaxies at these redshifts. 

\subsection{Accreted Black Hole Mass Density}

From the stacked upper limits to the X-ray luminosity we can derive the accreted black hole mass density in these galaxies. In order to do this we follow the procedure outlined in the
supplementary information section presented by \citet{treister11}. Briefly, we base our calculation on the so-called ``Soltan argument'' \citep{soltan82}, from which we have that the integrated
black hole mass density is given by:

\begin{equation}
\rho_{BH}(z)=\int^\infty_z\frac{dt}{dz}{dz}\int^\infty_0\frac{1-\epsilon}{\epsilon c^2}L_{bol}\Psi(L,z)dL,
\end{equation}

where

\begin{equation}
L_{bol}=k_{corr} L_X
\end{equation}

and $k_{corr}$ is the bolometric correction for the rest-frame hard X-ray band. For simplicity, following \citet{treister11}, we assume that $k_{corr}$=25 independent of luminosity and redshift.
Then, assuming that the AGN luminosity function does not evolve significantly at $z$$>$6 we obtain that:

\begin{equation}
\rho_{BH}(z)=\frac{(1-\epsilon)k_{\rm corr}}{\epsilon c^2}\int^\infty_z\frac{dt}{dz}{dz}\int^\infty_0L_X\Psi(L)dL.
\end{equation}

Here, the second integral on the right side can be determined directly from the observed integrated AGN emissivity.
Further assuming a constant radiation efficiency $\epsilon$=0.1 we obtain the upper limits for the accreted black hole mass
density at $z$$\sim$6,7 and 8 reported in Table 1. In comparison, following the same procedure but using the upper limits 
in the rest-frame hard band and number of sources reported by \citet{basu-zych13}, we obtain BH mass densities of 990, 1142 and 
1263 M$_\odot$~Mpc$^{-3}$ at $z$$\sim$6,7 and 8 respectively. These values are slightly higher but fully consistent with our results.

\begin{figure}
\begin{center}
\includegraphics[width=0.5\textwidth]{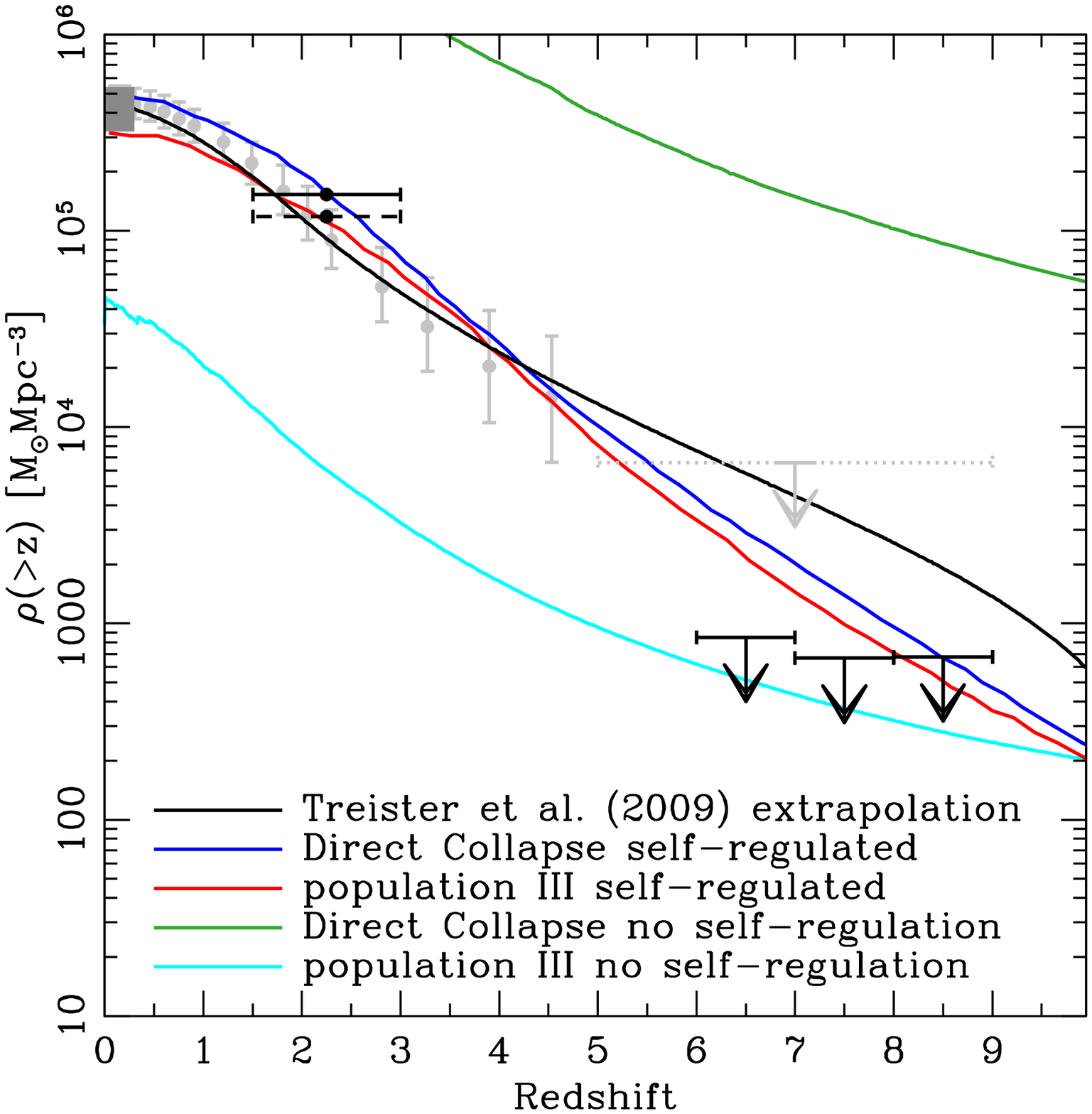}
\end{center}
\caption{Accreted BH mass density as a function of redshift. The {\it gray rectangle} shows the range of values allowed by observations of 
$z$$\simeq$0 galaxies \citep{shankar09}, while the {\it filled circles} show the observations of BH accreted mass density traced by AGN activity 
in the 0$<$$z$$<$5 range compiled by \citet{hopkins07} and measured by \citet{treister09c} at $z$$\sim$2. At higher redshifts, we show the CXRB integral 
constraint derived by \citet[][upper limit with dotted error bars]{salvaterra12} and the upper limits at $z$=6-9 obtained in this work. The {\it red} and
{\it blue} lines show the predicted BH mass density if self-regulation, as described in section 3.2, is incorporated, for Pop III and direct collapse BH seeds respectively.
The cyan and green lines show the results from these models if no regulation is assumed.}
\label{bhmass_w_z}
\end{figure}

In Fig.~\ref{bhmass_w_z} we show the density of accreted mass onto SMBHs as a function of redshift. In addition to our
upper limits at 6$<$$z$$<$8 , we include the constrains derived by \citet{salvaterra12} using the unresolved fraction
of the cosmic X-ray background (CXRB), which are less restrictive than our upper limits. However, in contrast to our work,
they are independent of the completeness of the galaxy sample studied and correspond to an integral constraint. But, as
discussed in \citet{salvaterra12}, they are affected by degeneracies in the assumed CXRB model, in particular with
low-luminosity sources at lower redshifts. Furthermore, as argued by \citet{treister09b}, the uncertainty in the measurements of the
CXRB and discrepancies between those derived by different missions can be $\sim$10\% of the total value, which is orders
of magnitude larger than the signal expected to be measured. 

In Figure~\ref{bhmass_w_z} we compare these upper limits with the predictions from the models of \citet{volonteri10a}. In these models, two ``seed'' formation 
models are considered: those deriving from population-III star remnants (Pop III), and from direct collapse models (D.C.). Independent of the seed mass, in this scheme
black holes grow primarily via accretion episodes triggered by galaxy mergers. This growth can be either self-regulated or un-regulated. Details of the implementation of
this self-regulation can be found in the supplementary section of \citet{treister11}. In a nutshell, in the self-regulated model, each black hole accretes an amount of mass, corresponding 
to 90\% of the mass predicted by the local $M_{\rm BH}-\sigma$ relation \citep{gultekin09}, while in the unregulated  mode the mass is simply doubled during each accretion episode.
For this set of models, the only one that satisfies the $z$$>$6 upper limits assumes light black hole seeds (Pop III) and no self-regulation. However, this model fails to account for the observed BH mass density 
at lower redshifts.

\begin{figure}
\begin{center}
\includegraphics[width=0.5\textwidth]{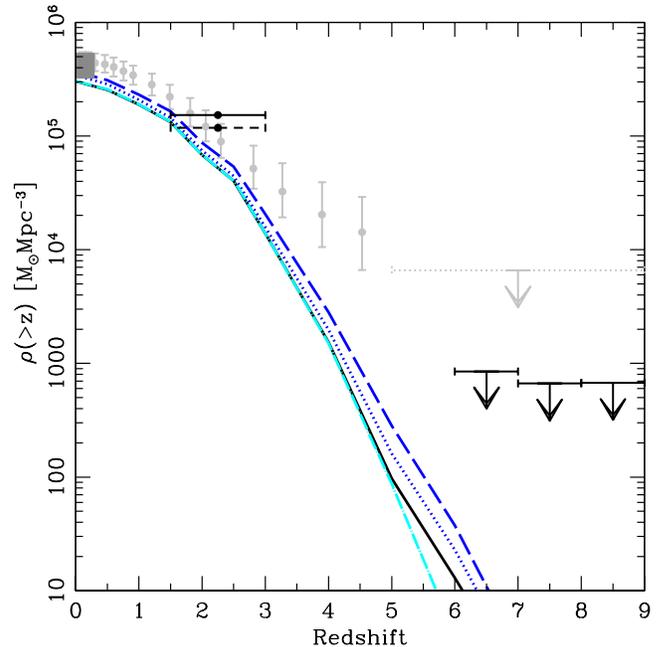}
\end{center}
\caption{Accreted BH mass density as a function of redshift, comparing with the models of \citet{bonoli12}. Symbols for the observational data are the same as in Fig.~\ref{bhmass_w_z}.
The dot-dashed cyan line only considers light (Pop III) BH seeds, while the solid black line and the blue dotted line include both Pop III and direct collapse seeds. The dashed blue line also considers
both seed types, but in this case massive seeds can also form in 1:10 mergers. For these models, marginal agreement with observations is obtained at $z$$<$3, while they are
at the same time consistent with our upper limits at $z$$>$6.}
\label{bhmass_w_z_bonoli}
\end{figure}

Other models, \citep[e.g.,][]{booth09,dubois12} predict a much steeper decline in BH mass density at high redshifts and thus are consistent with our observed upper limits. As an example,
in Fig.~\ref{bhmass_w_z_bonoli}, we compare our observational results with the predictions from the models of \citet{bonoli12}, which incorporate both light (Pop III) and massive (direct collapse)
BH seeds. These models assume that a massive, direct collapse, seed form whenever there is a major merger ($<$1:3 mass ratio) of massive galaxies which do not already contain a massive BH,
as suggested by \citep{mayer10}, although this scenario was later questioned by \citet{ferrara13}. Further BH growth is then triggered by both major and minor mergers. The main differences between the models compared here to observations, which can explain the differences in BH mass density at high redshift, are the formation epoch of black hole seeds that extends up to nearly $z$$\sim$0 in the \citet{bonoli12} models, 
the assumed average accretion rate, linked to the triggering methods, and their redshift dependence. As can be seen in Fig.~\ref{bhmass_w_z_bonoli}, while these models are consistent with our observed upper limits at $z$$>$6, they are only marginally consistent with lower redshift constraints. Furthermore,
it is important to note that these models do not include BH growth by secular processes, which can represent a significant fraction of the total BH accretion \citep{treister12, bellovary13}, and that the strong decrease in 
mass density at high redshifts can be due to the limited resolution of the simulations. In summary, we can conclude that spanning the whole range of BH growth, from the most luminous quasars which require massive 
BHs and very high near-Eddington accretion rates, to our upper limits at $z$$>$6, which suggest accretion levels lower than $\sim$10\% Eddington, is very difficult to track in simulations and requires strong redshift 
evolution in most scaling relations \citep{volonteri11}.

\begin{figure}
\begin{center}
\includegraphics[width=0.5\textwidth]{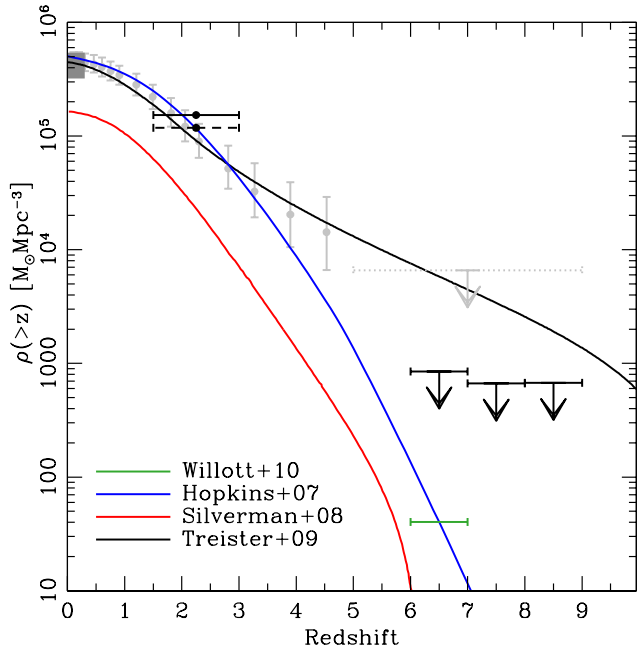}
\end{center}
\caption{Comparison between the observed accreted mass density in SMBHs and expectations from AGN LFs. Observed symbols are
the same as in Fig.~\ref{bhmass_w_z}. The {\it red} and {\it blue}  lines show the values inferred
from the LFs of \citet{silverman08b} and \citet{hopkins07} respectively, while the point at $z$=6-7 was obtained from the quasar LF 
of \citet{willott10a}. The {\it black line} assumes the hard X-ray AGN LF of \citet{ueda03}, as modified by \citet{treister09b}.}
\label{bhmass_w_z_lf}
\end{figure}

Comparing with extrapolations of AGN LFs, as shown in Figure~\ref{bhmass_w_z_lf}, the value reported by \citet{willott10a} --- based on observations of high-$z$ 
optical quasars --- is consistent with the upper limits derived here. Similarly, the bolometric luminosity function of \citet{hopkins07} provides a 
good description of the observed accreted BH mass density at $z$$<$4 and is consistent with our upper limits. In contrast,
while the prediction of \citet{treister09b}, based on an extrapolation of the \citet{ueda03} AGN LF, provides the best description of the observational data up to $z$$\sim$5 is clearly inconsistent 
with the 6$<$$z$$<$9 upper limits derived in this work, and in marginal agreement with the constraints of \citet{salvaterra12}. 

It is important to note that the tension between the upper limits reported here and at least some of the extrapolations of the AGN LF to high redshifts
can be alleviated if uncertainties in the former are considered. For example, a constant radiation efficiency was assumed, while it is possible that most of the
BH growth is radiatively inefficient. Similarly, if most of the black hole growth at high redshift is heavily obscured, it would change our assumed bolometric 
corrections, thus moving the reported upper limits towards higher accreted black hole mass densities. These effects and others are discussed in more detail
in the following section.

\subsection{Where are the Growing Black Holes?}

Our observations thus pose a puzzle: the X-ray stack of a sample of star-forming high-redshift dropout galaxies shows rather feeble if no signs of BH accretion,  despite a favorable environment for accretion - a gas-rich galaxy.  We now reflect on what this means both in the context of existing 
models  \citep[e.g.,][]{volonteri10a} and extrapolations of the AGN LF \citep[e.g.,][]{treister09b}. We discuss a number of possible explanations, which are neither exhaustive nor mutually exclusive. We note the there is a crucial distinction between occupation fraction for SMBHs and AGN fraction. The occupation fraction is a measure of whether a galaxy or halo is seeded with a BH regardless of whether it is actively accreting. If it is actively accreting, its classified as an AGN and contributes to the estimate of the AGN fraction. For example, even with a high occupation fraction, it is possible that only in a small subset of the galaxies studied here the BH is actively accreting and growing, thus revealing itself as an AGN.

The first possibility to consider is that the galaxies included in our samples do not contain SMBHs or only a small fraction of them do. If this is the case, this is telling us about the efficiency of seed formation. Indeed, as described by e.g., \citet{menou01}, it is theoretically possible and permissible that at high redshifts only a small fraction of the galaxies actually harbor a SMBH. For example, \citet{menou01} showed that with a BH occupation fraction as low as $\sim$10\% at $z$=5 there are ample seeds for all the galaxies in the local Universe to harbor a central SMBH. This is explained by the sequence
of mergers that dark matter halos will experience across cosmic history that will populate black holes into galactic nuclei that are initially bereft of them. What fraction of occupied  galaxies host actively accreting ones is a further question. If indeed less than $\sim$1 in 10 of the galaxies studied here actually contain a SMBH and perhaps only a fraction of them are actively growing, our X-ray stacking procedure will not be sensitive enough to detect them. This could clearly explain and account for our observational results. 

However, it is possible that our sample of LBGs contain SMBHs, but they are not actually growing. Indeed, at lower redshifts, $z$$\sim$3, the fraction of AGN in Lyman break galaxies is relatively low, $\sim$3\% \citep{nandra02,laird06,hainline12}. Likewise, in the  recent study of \citet{cowie12}, samples of Lyman break galaxies at $z$$\sim$3,4 and 5 were  not detected in sensitive X-ray stacks, thus suggesting a low AGN fraction for our sample of LBGs at $z$$>$6 if this trend were to hold at higher redshifts as well. Therefore, while the occupation fraction might be high, the AGN fraction appears to be very low for this sample. 

As a caveat, there are significant differences between those lower redshift samples and the $z$$>$6 Lyman break galaxies studied in our work here. For example, high redshift LBGs have significantly lower stellar masses and similar or higher star formation rates (by no more than a factor of two difference between z $\sim$3 and z $\sim$4) \citep{verma07} and \citep{reddy12}, and thus much higher sSFRs, indicative of higher gas fractions. 
In recent studies, \citet{de-barros12} and \citet{stark13} report a significant increase in the sSFRs observed in LBGs with redshifts. In addition, at $z$$<$3 it has been shown that the AGN fraction is a strong function
of the stellar mass of the host galaxy \citep[][and references therein]{xue10,mullaney12}, ranging from a few percent at $\sim$10$^9$M$_\odot$ to $>$20\% at $>$10$^{11}$M$_\odot$, thus suggesting a low AGN fraction for
our sample of LBGs at $z$$>$6. However, while our galaxies, at stellar masses $\sim$10$^9$M$_\odot$, will be amongst the less massive galaxies at $z$$<$3, they are some of the most massive ones at $z$$>$6, as even 
$\sim$10$^{10}$M$_\odot$ galaxies are very rare \citep{gonzalez11}. These differences strongly suggest that while the AGN fraction in relatively low-$z$ LBGs is low, the expectation is for a higher fraction in 
higher redshift LBG populations.

The very low accretion rates estimated in these galaxies is particularly puzzling as the host galaxies are known to have very high 
specific SFRs (sSFRs) of $\sim$2-20 Gyr$^{-1}$ \citep{gonzalez10}. This means that the galaxies that make up our X-ray stacks 
are efficiently converting gas to stars, which in turn corresponds to high gas fractions. These galaxies, appear however, unable to fuel growth of a central SMBH.  One plausible explanation for our results is that only a small fraction of these drop-out galaxies actually harbor a central BH. But if they do indeed harbor BHs at their centers, why are  they not accreting with a significant duty cycle?  On the one hand, our results mimic the findings of \cite{laird06} X-ray analysis of $z\sim 3$ LBGs, wherein they report no evidence that LBGs, which are certainly galaxies in which active star formation is occurring, are also preferentially active in nuclear black hole accretion. On the other hand, the non-LBG population at lower redshift ($z$$\sim$3), with comparable sSFRs to these higher redshift sources do appear to have high observed  AGN fractions, albeit at higher stellar masses, \citep{mullaney12}. 

The globally averaged star formation rate density as a function of redshift appears to track the accretion rate density onto luminous quasars rather well, and this has led us to believe that star formation and black hole growth occur in tandem, at least from a statistical point of view \citep{merloni04,zheng09}. However, this definitely does not imply that this concordance occurs in every galaxy. LBGs might just not be the sites that harbor the most actively growing black holes at  $z$$>$6. By selection, our galaxy sample is composed of the most massive galaxies at these redshifts, with  stellar masses $\sim$10$^9$M$_\odot$, and are essentially dust-free \citep[e.g.,][]{bouwens10c,wilkins11}. Thus, it is possible that our stack does not contain the population of galaxies whose BHs are actively growing at 6$<$$z$$<$8, if this growth is restricted to dustier and/or less massive galaxies. Such galaxies will be below the detection threshold for even the deepest optical/near-IR surveys carried out by large ground-based telescopes or the HST. This implies either that only relatively small black holes are growing in the early Universe or that it is possible at high redshifts for small galaxies to contain substantial central black holes. This possibility of obese BH galaxies has been recently explored by \citet{agarwal13}. 

This scenario indicates that while it appears that the globally averaged SFR and BH accretion rates track each other in the Universe by and large,  these properties are not tightly coupled for all individual sub-samples/sub-populations of galaxies. Hence, co-evolution may not occur for every galaxy at the same time. For instance, in most optically-detected quasars the SMBH accretion rate is typically much higher than $10^{-3} \times$ the SFR \citep[cf. Fig.~2 in][]{willott13}, while X-ray stacking of $z$$\sim$2 star forming galaxies suggests the SMBH accretion rate is at the level $10^{-3} \times$ the SFR, the scaling needed for co-evolution of the SMBH mass with stellar mass \citep{mullaney12}. Clearly the relative timescales of SMBH growth and SFR \citep {netzer09} are a key determinant in the establishment of the correlations between SMBHs and host galaxies, as well as in the interpretation of observational results.

Alternatively, it is also possible that these samples contain a significant fraction of low-$z$ interlopers. The most likely interlopers for high-z galaxy samples include reddened and/or old galaxies 
at $z$$\sim$1-2, low mass stars and spurious or transient sources. All of them have lower X-ray fluxes than average AGN and therefore would artificially decrease the signal in our stacks.
The contamination fraction in $z$$>$6 galaxy samples have been extensively debated in the past, as shown by example in Appendix D of \citet{bouwens06}. These results, and others, indicate that 
typical contamination levels are $\sim$10\% or lower \citep[e.g.,][]{bouwens11a}. If this is indeed the case, the existence of these relatively insignificant fraction of low redshift interlopers cannot significantly 
change our results or explain the lack of an X-ray detection. However, it is possible that these contamination levels have been significantly underestimated, given that the contribution of more exotic galaxies,
such as those with extreme emission lines \citep{brammer13} cannot be properly accounted for. 

As proposed by \citet{treister11}, it is possible that a large fraction of the emission due to accretion onto SMBHs in the early Universe is obscured 
by large amounts of gas and dust. Our more restrictive constraints here were obtained from the observed-frame soft Chandra band, 0.5-2 keV, which at these high redshifts corresponds to rest-frame
energies of $\sim$2-10 keV, i.e., the observed-frame hard Chandra band. Therefore, for these observations to
be significantly affected by obscuration would require extremely high levels of obscuration, up to Compton-thick column densities,
$N_H$$\sim$10$^{24}$cm$^{-2}$. While an increased contribution at high redshift of such heavy obscurations is certainly possible \citep[e.g.,][]{moretti12}, the observations in the observed-frame
hard band trace rest-frame energies of $\sim$30~keV, at the peak of the AGN X-ray emission, even for heavily obscured sources.
These upper limits, which are roughly $\sim$10 times higher than those obtained in the observed-frame soft Chandra band, while 
would be in marginal agreement with most existing models and expectations, will still generate tension. The lack of any mildly obscured and unobscured AGN at this redshift 
also raises puzzling questions regarding Unification; though perhaps the explanation lies in the high gas density expected in high-sSFR, compact galaxies - high 
column densities in all (4$\pi$) directions. Indeed, at lower redshifts \citet{xue12} found that most of the low luminosity AGN at similarly low host galaxy stellar masses show 
evidence for significant obscuration.

Finally, we are implicitly assuming here that black hole growth is due to a radiation-efficient matter accretion \citep[i.e., the][argument]{soltan82}. While this is 
certainly true in general \citep[e.g.,][]{yu02,marconi04}, it is possible that is not the case in the early Universe. Specifically, and as argued by \cite{shapiro05} and \citet{petri12} at $z$$>$6 BH growth due to mergers,
might dominate over accretion processes. If this is indeed the case, BH growth may not be accompanied by luminous 
emission, or at least not electromagnetic radiation, and thus our X-ray observations are unable to detect it.

\subsection{Black Hole Masses}

The lack of detection of luminous, individually detected, AGN at $z$$>$6 in deep X-ray surveys, together with the upper limits reported here pose interesting and strong limits on the early growth and formation 
mechanisms for SMBHs. For example, assuming a canonical 10\% bolometric correction for hard X-ray emission \citep[e.g.,][]{treister09b} and accretion at the Eddington limit implies
average BH masses smaller than 2.7$\times$10$^{4}$M$_\odot$ for the $z$$\sim$6 sample (and lower than $\sim$3$\times$10$^5$M$_\odot$ for a more typical 10\% Eddington
ratio), for an AGN fraction of 100\% (if a lower AGN fraction is assumed instead, BH masses should be scaled upwards accordingly. For example, they will be smaller than $\sim$3$\times$10$^6$M$_\odot$ for a 10\% AGN fraction at a 10\% Eddington ratio).  Furthermore, the 
luminous quasars detected by optical surveys at $z$$>$6 do contain massive BHs ($>$10$^8$M$_\odot$) that appear to be accreting near their 
Eddington limits \citep{willott10b}. If the typical galaxy at those redshifts had a smaller BH (scaled down version from the most luminous quasars) growing at similar Eddington ratios or even 
lower by an order of magnitude, we would have detected them in our study (once the caveats presented in \S 3.2 are taken into account). 

In contrast, many models used to explain the formation and evolution of SMBHs assume large Eddington ratios, $\ga$30\%, thus
resulting in relatively massive BHs growing rapidly in the early Universe. For example, the models of \citet{volonteri10a},
in which the accretion is driven by merger-triggered episodes, that tend to bring the resulting BH into the observed M-$\sigma$ correlation, predict that by $z$$\sim$6 the average masses of the actively-growing BHs should be in the $\sim$10$^6$M$_\odot$ range, in marginal agreement with our observations, if relatively low accretion rates and AGN fractions are assumed. A key point to note here about the models is that they hinge on the circular velocity (and therefore halo mass) as the lever for black hole seed masses and the regulation of growth during accretion episodes. In fact, in all current models it is assumed that the  accreted mass is proportional to some power of the circular velocity. We caution that it is unclear what the halo masses are for these drop-out galaxies, what we have are only estimates of their average stellar masses. Most BH growth models provide scaling relationships specifically between halo mass and black hole mass and not stellar mass, so direct comparison is complicated by the need to make additional assumptions about the efficiency of star formation in the early Universe. 

Given the observed correlation between BH mass and both bulge \citep{haring04} and total stellar mass \citep{jahnke09} for nearby galaxies, we can estimate the BH mass corresponding to the galaxies in our sample, assuming of course that this correlation holds at
high redshifts. Stellar masses were provided for the F12 sample, as derived from spectral fitting. The average stellar mass
for the $z$$\sim$6 galaxy sample is 10$^{9}$M$_\odot$ \citep[see also][]{curtis-lake13}, which implies using the \citet{jahnke09} relation that typical BH masses
are expected to be $\sim$10$^6$M$_\odot$. This assumes no redshift evolution in the M$_{BH}$-M$_*$ relation, consistent with the findings up to $z$$\sim$1 of e.g., \citet{cisternas11a}. Such BH masses are still consistent with the upper limits measured here, provided that both the AGN fraction and the Eddington accretion rates are $\sim$10\% or lower. The growth rate of a BH scales as
\begin{equation}
M(t)=M_0\times\exp(f_{edd}\times\Delta t\times(1-\epsilon)/(\epsilon\times0.45 Gyr)),
\end{equation}

where $\Delta$t is the time allowed for growth, and $f_{edd}$ is the Eddington ratio, $\sim$0.1 in the example above. An AGN fraction of 10\% 
translates into a 10\% duty cycle, i.e., $\Delta$t is $<$10\% of the Hubble time at the redshift of interest, of order 1 Gyr at $z$$\sim$6. Since with 
these figures the term in the exponent is very small, 0.2 or less, this implies that in this picture BHs do not gain significant mass in the early
stages of the Universe.

\section{Conclusions}

We present here the X-ray properties of samples of $z$$\sim$6,7 and 8 galaxy candidates in the CDF-S. None of these galaxies are detected in X-ray, either individually or collectively via  stacking. This non-detection via the consequent upper limit on the accreted mass density of $<$1000~M$_\odot$Mpc$^{-3}$, offers the most stringent constraints on black hole growth in the early Universe. This is particularly surprising, as our X-ray stacking observations are sensitive enough to detect even moderate amounts of accretion in relatively small black holes. Furthermore, such low accretion levels contradict the predictions of several black hole formation and  evolution models and the expectations based on extrapolations of existing AGN luminosity functions. 

Explaining these results requires that these high-redshift dropout galaxies, which are now routinely found and studied by HST and large ground-based telescopes, (A) do not contain SMBHs, or (B) if they contain black holes, then these are not growing, or (C) the black hole growth if occurring is heavily obscured and/or not radiating efficiently in X-rays. If BHs are present in these galaxies, we return to the question of why they are not accreting, in particular since they appear to have significant amounts of gas, given their high specific star formation rates. If they do not contain BHs, then our results have new implications for BH seed formation mechanisms; namely, that normal star-forming galaxies  at 6$<$$z$$<$8 are not forming/growing BH seeds at their centers. This may indicate that seed formation and growth in the general galaxy population can be delayed in some cases until much lower redshifts, as suggested by \citet{volonteri10} and more recently by \citet{bonoli12}. A particularly remarkable example of such systems was discovered and reported by \citet{schawinski11a} at $z$=1.35. Our results strongly suggest that the individual sites where the bulk of the stars form at 6$<$$z$$<$8 may not be the sites that harbor the most massive black holes.

Interestingly many similarities exist between our result and the conclusions drawn by \cite{willott10b}, based on the comparison between the SMBH and galaxy mass functions at $z$$\sim$6, that most galaxies at that time formed their stars much more rapidly than their SMBHs grew or that SMBH seeding is inefficient. This conclusion does not of course apply to the most luminous quasars at that epoch, where an increase of the ratio between SMBH mass and galaxy mass is in fact observed. Therefore, the most massive SMBHs in the most massive galaxies/halos are actively accreting and are on or above the correlation with their hosts \citep[e.g.,][and references therein]{wang10}, while at lower galaxy mass either many galaxies do not have SMBHs, or these SMBHs are much less massive than expected, or they are not accreting or obscured. Therefore, there could be significant variations in the strength of the correlation between BH growth and star formation in individual galaxies, or its dependence on galaxy mass \citep{volonteri11}.

\acknowledgements
We thank the anonymous referee for a careful and thoughtful reading and several useful suggestions, and Dan Coe and Rychard Bouwens 
for useful discussions on the HUDF data. Support for the work of ET was provided by the Center of Excellence in Astrophysics and 
Associated Technologies (PFB 06), by the FONDECYT regular grant 1120061 and by the Anillo project ACT1101. KS gratefully 
acknowledges support from Swiss National Science Foundation Grant PP00P2\_138979/1. PN acknowledges support from the NSF TCAN 
program via grant AST-1332858.


\begin{deluxetable}{lcccccccc}
\tablecolumns{9}
\tablewidth{0pc}
\tabletypesize{\scriptsize}
\tablecaption{Stacking Results}
\tablehead{
\colhead{Redshift} & \multicolumn{4}{c}{Net counts\tablenotemark{a}} & \colhead{Count Rate\tablenotemark{b}} & \colhead{Flux\tablenotemark{b}} & \colhead{Lum\tablenotemark{b}} & \colhead{BH Mass\tablenotemark{b,c}}\\
\colhead{} & B06 & B11 & F12 & Combined & \colhead{ (3$\sigma$) [s$^{-1}$]} & \colhead{[erg cm$^{-2}$ s$^{-1}$]} & \colhead{[erg s$^{-1}$]} & \colhead{[M$_\odot$Mpc$^{-3}$]}\\
\hline\\
\multicolumn{9}{c}{{\bf Soft Band (0.5-2 keV)}}}
\startdata
$z$$\sim$6 & -3.4$\pm$6.2 & --- & -3.6$\pm$4.7 & -4.0$\pm$6.5 (272)\tablenotemark{d} & $<$8.9$\times$10$^{-8}$ & $<$6.5$\times$10$^{-19}$ & $<$2.6$\times$10$^{41}$ & $<$851\\
$z$$\sim$7 & --- & 0.7$\pm$1.4  & -0.6$\pm$2.5 & -0.4$\pm$2.6 (46)\tablenotemark{d} & $<$1.9$\times$10$^{-7}$ & $<$1.2$\times$10$^{-18}$ & $<$6.8$\times$10$^{41}$ & $<$666\\
$z$$\sim$8 & --- & 1.6$\pm$1.7  & 0.7$\pm$1.9 & 1.9$\pm$2.1(23)\tablenotemark{d} & $<$3.0$\times$10$^{-7}$ & $<$2.0$\times$10$^{-18}$ & $<$1.5$\times$10$^{42}$ & $<$674\\
\cutinhead{{\bf Hard Band (2-8 keV)}}
$z$$\sim$6 & -6.3$\pm$9.1 & --- & -3.3$\pm$6.7 & -9.1$\pm$9.4 & $<$1.7$\times$10$^{-7}$ & $<$3.4$\times$10$^{-18}$ & $<$1.6$\times$10$^{42}$ & $<$4750\\
$z$$\sim$7 & --- & 0.2$\pm$2.4  & 1.8$\pm$4.1 & 1.5$\pm$4.1 & $<$3.9$\times$10$^{-7}$ & $<$7.9$\times$10$^{-18}$ &$<$5.3$\times$10$^{42}$ & $<$4704\\
$z$$\sim$8 & --- & -4.7$\pm$2.1  & -0.4$\pm$2.6 & -1.8$\pm$2.8 & $<$5.4$\times$10$^{-7}$ & $<$1.1$\times$10$^{-17}$ & $<$9.8$\times$10$^{42}$ & $<$4346\\
\enddata
\tablenotetext{a}{Background subtracted, stacked, counts}
\tablenotetext{b}{For the combined sample}
\tablenotetext{c}{Accreted Black Hole mass density}
\tablenotetext{d}{Number of stacked sources in the combined sample}
\label{stack_results}
\end{deluxetable}

\end{document}